\documentclass[pra,nofootinbib]{revtex4}

\usepackage{graphicx}
\usepackage{amssymb}
\usepackage{amsmath}
\usepackage{amsfonts}
\usepackage{epstopdf}
\usepackage{epsfig}
\usepackage{wrapfig}

\begin{document}
\title{Information is not physical}

\author{Robert Alicki,}
\affiliation{Institute of Theoretical Physics and Astrophysics, University of
Gda\'nsk, Poland }

\begin{abstract}
The standard relations between information theory and thermodynamics are challenged. The Szilard engine is revisited and the bound proposed by Landauer
 is replaced by a different one which includes errors in information processing. Instead of equivalence, complementarity of information and thermodynamical entropy is advocated.
Finally, the relations between error correction and self-replication of states which can carry information are discussed.
\end{abstract} 
\maketitle

\section{Introduction}

The title of this essay is obviously polemic with the title of the first Section -\emph{Information is physical} - of the famous Rolf Landauer's paper \emph{The physical nature of information} \cite{Landauer:1996}. This paper was quite influential and stimulated, among the others, the modern advances in the study of relations between information theory and thermodynamics. This new wave of research is parallel  to the rapid development of ``quantum information'' and ``quantum computing'' (although Landauer himself was quite skeptical about the later). The aim of my note is to challenge the popular opinions concerning relations between physics and information theory (see \cite{Stanford} for references to other critical contributions). In particular, I would like to present the following thesis and support at least some of them by suitable  models :\\

I) Information $I$, measured for simplicity in \emph{nats} ($\log_2 e \simeq1.443$ bits), \textbf{is a disembodied abstract entity independent of its physical carrier}. In particular, multiplying $I$ by the Boltzmann constant $k_B$ or thermal energy  $k_B T$ does not yield thermodynamical entropy or meaningful energy which should be taken into account in the energy-entropy balance related to the First and Second Law of Thermodynamics. The direct interpretation of the title of the paper is  that the quantities like  $k_B\times I$ or $k_B T\times I$ possess no physical meaning.  \\

II) It is true that, as Landauer wrote :\emph{ "[Information] is always tied to a physical representation. It is represented
by engraving on a stone tablet, a spin, a charge, a hole in a punched card, a mark on paper, or some
other equivalent. This ties the handling of information to all the possibilities and restrictions of our real
physical word, its laws of physics and its storehouse"}.\\
However, the legitimate questions concern the physical properties of \textbf{information carriers} like  "stone tablet, a spin, a charge, a hole in a punched card, a mark on paper", but not the information itself. The examples of  legitimate questions are the following:

a) What is the error of encoding,  readout and generally processing of information carried by specific physical systems in  given environments?\\

b) What is the life-time of information encoded in a given physical system in a given environment?\\

c) What is the minimal amount of work needed to change the encoded information? \\

d) What is the time needed to perform a single step of information processing?\\

e) What are the relations between the quantities of above?\\

III) To encode information one needs physical systems possessing a number of distinguishable and stable with respect to thermal and quantum noise states. The distinguishability of states can be quantified  in terms of their overlap, well-defined for classical and quantum systems. Only well-distinguishable states can be cloned with a high accuracy and  protected against noise for long times. Therefore, for example, gas of atoms may possess a well-defined entropy but does not encode any information\footnote{A s-f writer and philosopher Stanis\l aw Lem  recognized very well this difference in his book \emph{The Cyberiad} (1964) using \emph{reductio ad absurdum}. In one of the tales two constructors, Trurl and Klapaucius are captured by a space pirate who pillages and hoards information. To gain their freedom, the constructors build , a "Demon of the Second Kind" designed to interpret the movement of air molecules as information. Whenever the motion of the molecules adds up to something intelligible, the Demon transcribes it onto paper tape using a tiny diamond-tipped pen. The pirate underestimates the amount of information contained within the chaotic motion, and he is soon buried in a mountain of paper filled with useless information: all the words that rhyme with spinach, why fan-tailed fleas won't eat moss, the sizes of bedroom slippers available on the continent of Cob, how Kipling would have written the beginning of The Jungle Book 2, etc., etc.[http://www.ethanham.com/blog/2009/08/popcorn-robots.html]}.
\\

IV) Information is neither classical nor quantum, it is independent of the properties of physical systems used to its processing. For example, in the problem of factoring of integers the input and output is information and can be quantified in bits or nats. The question of efficiency, i.e. the amount of resources needed to perform this task with a given level of confidence, is in principle a physical question. Only because our present day computers are based on macroscopic elements, which are highly stable on human time-scale and are applied to moderately large inputs and outputs appropriate again to the human scale, we can still disregard, to a large extend, physical limitations and apply abstract complexity theory to efficiency problems.  \\

V) The importance of "information" in physics is highly overestimated in the recent times. One often says that all physical phenomena are instances of "information processing". Such general statement possess a prediction power and hence a scientific value rather low. On the contrary, all fundamental theories starting from classical mechanics and ending with quantum field theory  and its possible new versions (string theory, quantum gravity, etc.) can be formulated without reference to any kind of information. Even entropy in statistical mechanics need not information-theoretical interpretation and can be treated simply as the (averaged) logarithm of  the number of accessible states.\\

VI) ``Quantum Information'' is the name of a fashionable field of research and not a different kind of information. The use of strongly overlapping quantum states as information carriers seems to be rather counter-factual as such states becomes more and more fragile when the size of the system increases. In contrast to many claims there are no proofs of quantum fault-tolerance based on physically acceptable assumptions. However, one cannot exclude that the high speed of  operations achieved for microscopic systems may compensate the high probability of errors and lead to a construction of some useful special purpose quantum devices of moderate size. In this case efficiency becomes a physical problem what makes formal ``quantum complexity theory'' not very useful. Ultimately, the  probability of success for any quantum computer scales down exponentially with the size of the task and no ``probability amplification`` is possible \footnote{Again Stanis\l aw Lem addressed this question in one of his tales (1964). \emph{Trurl and Klapaucius were former pupils of the great Cerebron of Umptor, who for forty-seven years in the School of Higher Neantical Nillity expounded the General Theory of Dragons. Everyone knows that dragons don’t exist. But while this simplistic formulation may satisfy the layman, it does not suffice for the scientific mind. Indeed, the banality of existence has been so amply demonstrated, there is no need for us to discuss it any further here. The School of Higher Neantical Nillity is in fact wholly unconcerned with what does exist. The brilliant Cerebron, attacking the problem analytically, discovered three distinct kinds of dragon: the mythical, the chimerical, and the purely hypothetical. They were all, one might say, nonexistent, but each nonexisted in an entirely different way.
But it turns out that dragons are merely highly improbable (one would have to wait a good sixteen quintoquadrillion heptillion years). So Trurl invents a probability amplifier and, when the local draconic probability is suitably heightened, a dragon materializes. 
The existence of the probability amplifier, and that the proven potential existence for dragons has lowered the further probability threshold of dragons materializing, that soon there are dragons being reported all over the galaxy} [The Dragons of Probability by Stanislaw Lem (translated by Michael Kandel)].}.
\section{Szilard engine revisited}
The mother of all misconceptions concerning the relations between information theory and thermodynamics is the Szilard's explanation of the Maxwell demon paradox in 1929.  While Smoluchowski in 1912 \cite{Smoluchowski} gave a mechanistic interpretation of the paradox in terms of the trap-door model of Maxwell demon which is a subject of thermal fluctuations, Szilard \cite{Szilard} introduced the notion of measurement into the game. The measurement understood as acquiring a bit (or $\ln 2$ nats) of information immediately led to the interpretation of  $k_B \ln 2$ as thermodynamical entropy
and $k_BT\ln 2$ as the minimal work which should be invested by external sources. This idea initiated a never ending discussion  about the place where the external work must be invested in order to save the Second Law of Thermodynamics \cite{Landauer:1961, Bennett:2003, Sagawa:2012}. Various arguments presented by different authors were based essentially of the Second Law combined with the postulate of
additivity of entropy for the joint system of information carrier and heat bath. This postulate has been critically discussed in \cite{Alicki:2012}.
\par
The Szilard engine consists of a box, containing only a single gas particle, in thermal contact with a heat bath, and a partition.
The partition can be inserted into the box, dividing it into two equal volumes, and is also capable of sliding, frictionless, along the box to the left or to the right. Now one assumes that \textbf{if it is known which side the molecule is on} ($\ln 2$ nats of information), it is possible to connect up the partition to a pulley and extract $k_B T\ln 2$ work in a process which can be repeated. 
\par
However, inserting the partition reduces the entropy of the single particle gas from the (poorly defined in the classical case) value $S$ to $S' = S - k_B \ln 2$ and hence increases the free energy  of the gas from the value $F = U - TS$ to the value  $F' = U - TS' = F + k_B T\ln 2$. \\

\textbf{The well-defined excess  free energy $\Delta F = k_B T\ln 2$, which can be transformed into useful work, is objective and does not depend on anybody's knowledge  ``of which side the molecule is on''.}\\

Indeed, it is not difficult to design, within the same level of idealization, procedures of extracting work without knowing the position of the particle (see Fig. 1(A)(B)) \footnote{This possibility was noticed long time ago by Popper and Feyerabend \cite{Feyerabend} with a goal to reject the idea that statistical mechanical entropy was a subjective quantity. However, it was unclear  what implications this had. For example, Feyerabend claimed explicitly that the second law of thermodynamics is violated, while to Popper it indicated that the second law is only applicable to large systems (see \cite{Stanford} for more information and references).}.

\begin{figure}[tb]
    \centering
    \includegraphics[width=0.8\textwidth,angle=270]{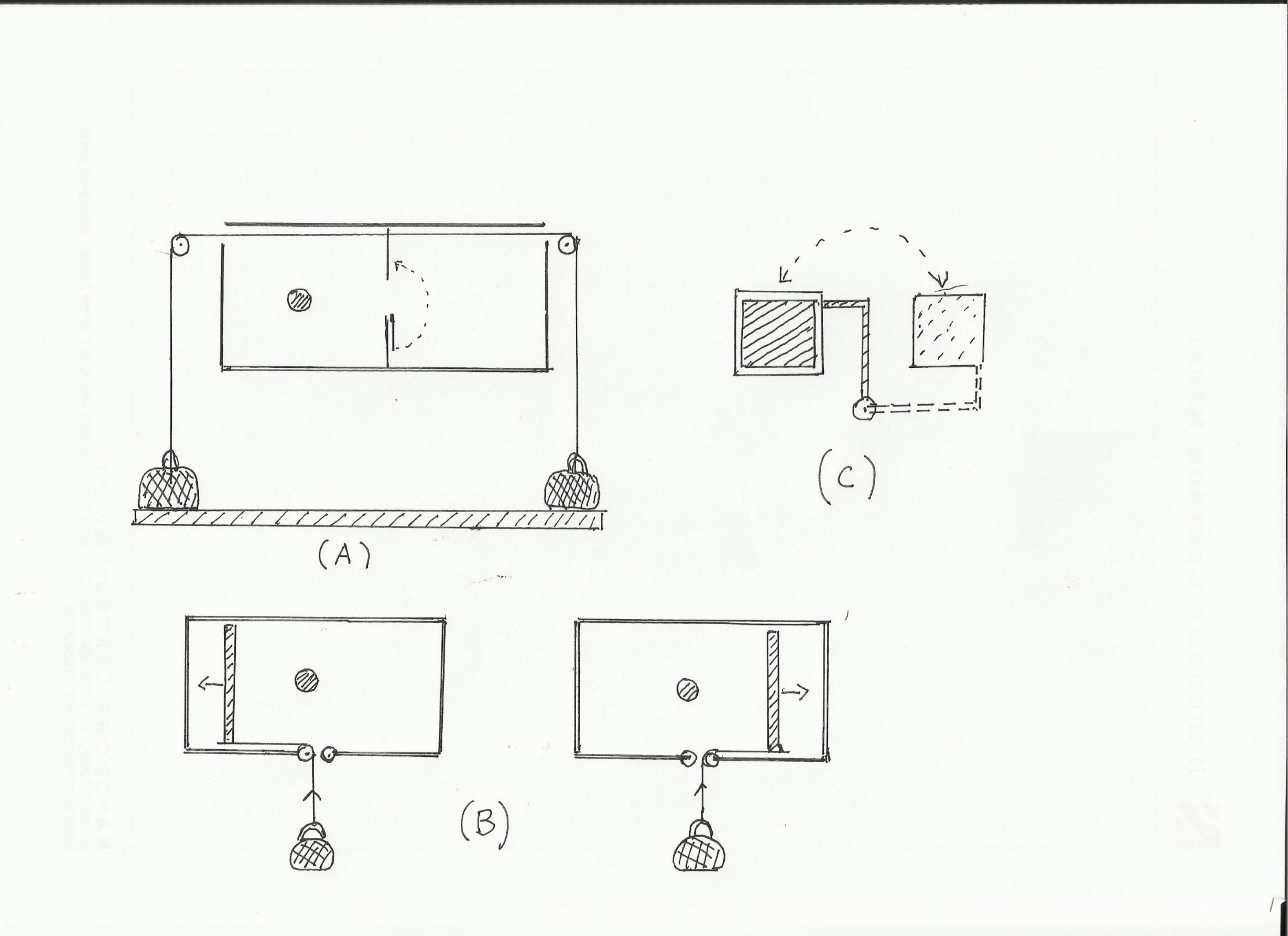}
    \caption{(A) Popper's design of the Szilard engine. (B) Another design which does not involve a measurement. (C) Partition as a switch with a potential barrier.}
    \label{nazwa1}
 \end{figure}

The missing work necessary to save the Second Law is hidden in the operation of the partition. Namely, the position of the partition inside the box must be stable with respect to thermal fluctuations and hence separated from the position ``partition outside the box'' by the energy barrier of the height $W >> k_BT$ (Fig.1C). Therefore to move the partition from one position to the other one needs at least several $k_B T$ of work to overcome the energy barrier\footnote{The erroneous assumption that: ``..in the ideal case, energy to place the block can be negligible..'' is behind the claim of ``experimental demonstration of information-to-energy conversion...'' in \cite{Toyabe}. In reality, ``placing a block'' (partition) costs at least several $k_BT$ of work.}. 
Friction is necessary to stabilize the partition in the new position and the supplied work is dissipated into environment. Summarizing, \textbf{ the operation of the Szilard engine can be easily explained without any reference to measurement and information}.
\section{A quantum model of a switch}
The basic element needed for operation of the Szilard engine is a partition which is an example of a \textbf{switch} - a system with (at least ) two distinguishable and stable states and a mechanism which changes one state to another in an irreversible process powered by the external source of work. One can think about all information carriers as ensembles of switches with given distinguishability of states and given stability with respect to external noise. 
\par
In this and next Sections I discuss a minimal quantum model of a switch and the thermodynamical bounds of its performance, and  I use this model to analyze a quantum version of the Szilard engine and its efficiency.
\par
The model, which was introduced and studied in details in the previous paper \cite{Alicki:2013} consists of two coupled quantum systems. The first one is microscopic, represented by a spin-$1/2$ and described by the standard Pauli matrices $\hat{\sigma}^k , k=1,2,3,\pm$. The second is the harmonic oscillator with  the canonical operators $\hat{a}, \hat{a}^{\dagger}$, which represent a pointer showing  the two positions of the switch. The corresponding states of the oscillator are assumed be be semi-classical an will be shown to be stable with respect to quantum and thermal noise. The  Hamiltonian of the total system contains a kind of ``ferromagnetic coupling'' which favorizes energetically particular position of oscillator which coincide with the spin direction along the $3$-axis. The switch Hamiltonian has the following form
\begin{equation}
\hat{H} =  \hbar\omega_0 \bigl[\hat{a}^{\dagger}\hat{a} - D(\hat{a}^{\dagger}+\hat{a})\hat{\sigma}^3 \bigr]
\label{ham_TLSQB}
\end{equation}
where, the dimensionless position of the oscillator is given by $\frac{1}{2}(\hat{a}^{\dagger}+\hat{a})$. In the following  spin eigen-states, oscillator coherent states and joint spin-oscillator states are denoted by
\begin{equation}
\hat{\sigma}^3 |\pm\rangle = \pm |\pm\rangle, \quad \hat{a}|\alpha\rangle = \alpha |\alpha\rangle, \quad \alpha\in\mathbf{Z},\quad |\mu;\alpha\rangle \equiv |\mu\rangle|\alpha\rangle, \quad \mu= \pm.
\label{coherent}
\end{equation}
Two degenerate ground states of the Hamiltonian \eqref{ham_TLSQB} are  product states of spin's ``up'' (``down'') states and corresponding oscillator states localized at the positions $\pm D$ and with averaged momenta equal to zero
\begin{equation}
|\Psi_{\pm}\rangle = |\pm ;\pm D\rangle .
\label{g_states}
\end{equation}
The information is encoded in the oscillator states $|\pm D\rangle $ only, which are not orthogonal.
The overlap ( transition probability ) between those coherent states  given by
\begin{equation}
\epsilon = |\langle D |- D\rangle|^2 = e^{-4D^2} .
\label{trans_0}
\end{equation}
can serve as an encoding error measure at zero temperature. One should notice that only  $\epsilon \in [0, 1/2)$ makes sense, for $\epsilon = 1/2$ oscillator states are completely indistinguishable. 
\par
In the next step  a weak interaction with a large quantum bath at the temperature  $T>0$ is added. In the simplest, minimal model this coupling is described by two interaction Hamiltonians  
\begin{eqnarray}
\label{Ham_int_1}
\hat{H}^{(o)}_{int} &=& (\hat{a} + \hat{a}^{\dagger}) \hat{F_o} ,\\
\hat{H}^{(1)}_{int} &=& \hat{\sigma}^1 \hat{F}_1 ,
\label{Ham_int_3}
\end{eqnarray}
where $F_{j}$ are independent environment observables. Using standard methods one can derive quantum Markovian master equations for the reduced density matrix of the switch. I am not going to repeat the detailed analysis of \cite{Alicki:2013} but restrict myself to the discussion of the final results.
\par
Omitting first the second Hamiltonian in \eqref{Ham_int_3} which describes  spin-flip one obtains the Markovian dynamics with  stationary states being mixtures
of two  of \emph{biased Gibbs states} 
\begin{equation}
\hat\rho^{(\pm)} =  \bigl(1-e^{-\hbar\omega_0/k_BT}\bigr)|\pm\rangle\langle\pm|\,e^{-\frac{\hbar\omega_0}{k_BT}(\hat{a}^{\dagger}\mp D)(\hat{a}\mp D)}.
\label{gibbs_biased}
\end{equation}
Any initial  state $\hat{\rho}$ tends asymptotically to the mixture $p_{+}\hat\rho^{(+)} + p_{-}\hat\rho^{(-)}$  with $p_{\pm}$ given by the initial probabilities of spin polarization. The corresponding  oscillator states are given by
\begin{equation}
\hat\rho^{(\pm)}_{\mathcal{P}} =  \bigl(1-e^{-\hbar\omega_0/k_BT}\bigr)e^{-\frac{\hbar\omega_0}{k_BT}(\hat{a}^{\dagger}\mp D)(\hat{a}\mp D)}.
\label{pointer_T}
\end{equation}
They are  well-localized and well-distinguishable (for $D>>1$) mixed Gaussian states with the overlap (identification, encoding, readout error)
\begin{equation}
\epsilon = \mathrm{Tr}\Bigl(\sqrt{\sqrt{\hat\rho_{\mathcal{P}}^{(+)}}\hat\rho_{\mathcal{P}}^{(-)}\sqrt{\hat\rho_{\mathcal{P}}^{(+)}}}\Bigr) =  \exp\Bigl\{-4 D^2\tanh\bigl(\frac{\hbar\omega_0}{2k_BT}\bigr)\Bigr\}
\label{trans_prob}
\end{equation}
The formula \eqref{trans_prob} can be rewritten in the more convenient form of the Boltzmann-like factor
\begin{equation}
\epsilon =  \exp\Bigl\{-\frac{W}{\Theta}\Bigr\} , \quad \Theta= \frac{\hbar\omega_0}{e^{\hbar\omega_0/k_BT} -1} + \frac{\hbar\omega_0}{2}.
\label{highT}
\end{equation}
The parameter $\Theta \equiv \Theta[T,\omega_0]$ is the average quantum oscillator energy at the temperature $T$ . $\Theta/k_B$ can be called  \emph{noise temperature} because it interpolates between temperature (for $T >> \omega_0$) and  (for $T<< \omega_0$) zero-point energy divided by $\hbar$, thus characterizing both, thermal and quantum components of the environmental noise. The energy parameter
\begin{equation}
W = \frac{1}{2}\Delta E = 2\hbar D^2\omega_0 . 
\label{work_0}
\end{equation}
is a half of energy amount $\Delta E= 4\hbar D^2\omega_0 $  necessary for spin-flip in one of the stable states $\hat\rho^{(\pm)}$ to obtain the excited biased Gibbs state
\begin{equation}
\hat\rho^{(\mp)}_*  = \hat{\sigma}^1\hat\rho^{(\pm)}\hat{\sigma}^1 = \bigl(1-e^{-\hbar\omega_0/k_BT}\bigr)|\mp\rangle\langle\mp|\,e^{-\frac{\hbar\omega_0}{k_BT}(\hat{a}^{\dagger}\mp D)(\hat{a}\mp D)}.
\label{gibbs_biased_ex}
\end{equation}
One can treat $W$ as a height of the energy barrier which separates states  $\hat\rho^{(\pm)}$ because the difference between the averaged energy of the state $\frac{1}{2}\hat\rho^{(\mp)}_* + \frac{1}{2}\hat\rho^{(\pm)} $ (placed ``halfway'' between $\hat\rho^{(\pm)}$ and $\hat\rho^{(\mp)}_*$)  and the averaged energy of $\hat\rho^{(\mp)}$ is exactly equal to $W$ (see Fig.2).

\begin{figure}[tb]
    \centering
    \includegraphics[width=0.8\textwidth,angle=270]{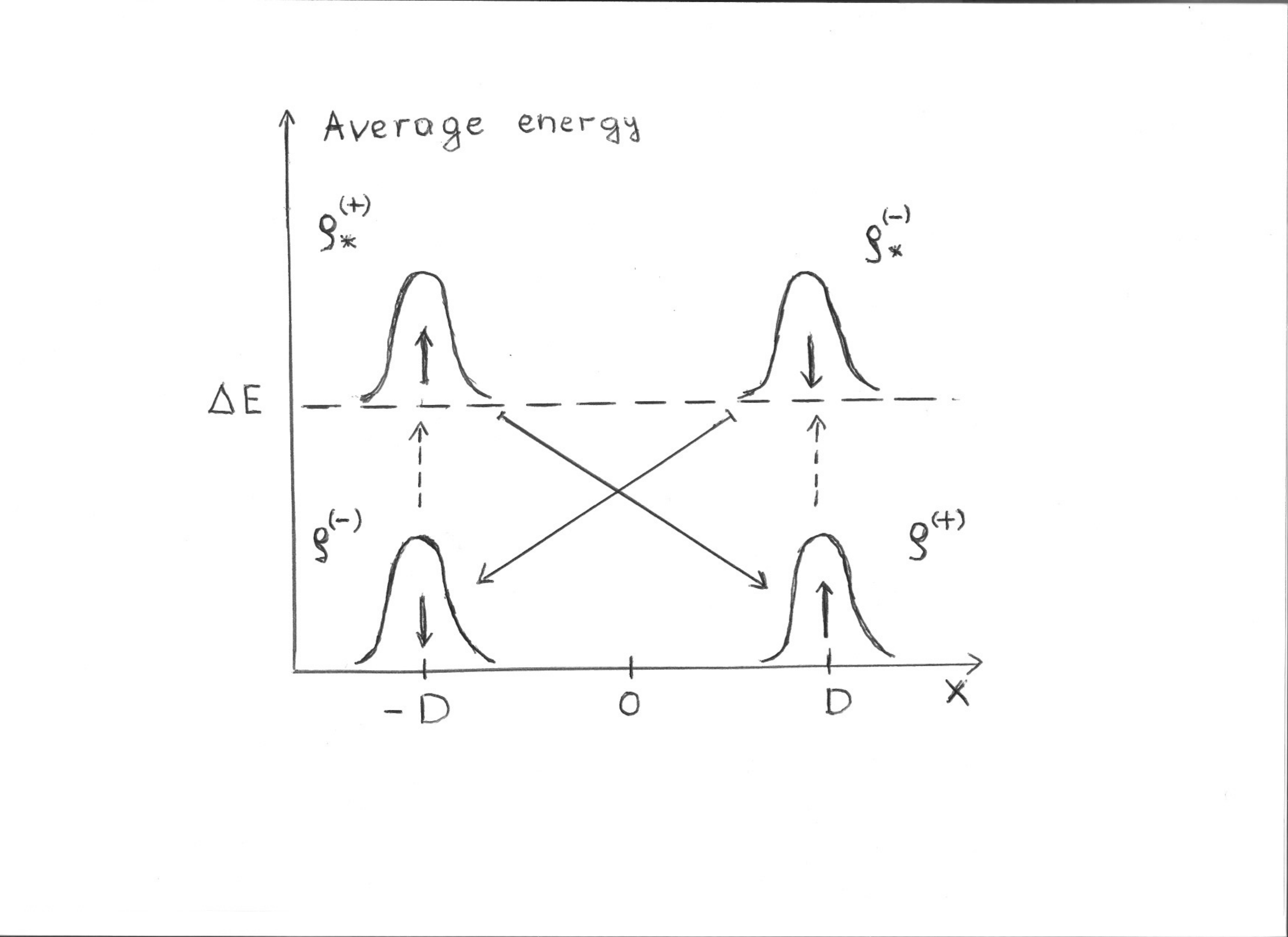}
    \caption{Stable states $\rho^{(\pm)}$ and their excitations $\rho^{(\pm)}_*$ . Gaussians depict localized oscillator states with arrows inside corresponding to spin states.  The long solid arrows represent dissipation routes, while the dashed ones first steps of tunneling process or the NOT gate.}
    \label{nazwa}
 \end{figure}

\par
The picture of protecting energy barrier $W$ suggests the mechanism which ensures stability of switch positions. One can expect that the life-time $\tau$ of information encoded in a switch interacting with a quantum heat bath should be given by the Kramers-like formula
\begin{equation}
\tau = \tau_0  \exp\Bigl\{\frac{W}{\Theta}\Bigr\}. 
\label{kramers}
\end{equation}
where $\tau_0$ is a life-time of unprotected information. Indeed it was shown \cite{Alicki:2013} for the discussed model of switch with the spin-flip process described by the second interaction Hamiltonian in \eqref{Ham_int_3}  that \eqref{kramers} gives a leading order approximation
for the life-time with $\tau_0$ being the life-time of the spin  decoupled from the oscillator, but interacting with the same environment.
\par
The operation of changing  position of the switch, which is called NOT gate can be executed using the following time-dependent Hamiltonian acting on the spin component%
\begin{equation}
\hat{H}_{NOT} = \hbar g(t)\hat{\sigma}^{1}
\label{ham_measure}
\end{equation}
where $g(t)$ is assumed to be a fast pulse concentrated around $t=0$ and satisfying $\int_{-\infty}^{\infty} g(t) dt = \pi$. For a fast enough pulse the initial state $\hat\rho^{(\pm)}$ is transformed to the excited state   $\hat\rho^{(\mp)}_*$ and then relaxes to $\hat\rho^{(\mp)}$ changing in an irreversible process switch position. The invested work is equal to $\Delta E = 2W = 4\hbar\omega_0^2 D^2$ and is completely dissipated during the relaxation process to the new position (Fig.2).
\section{Thermodynamics of information protection}
Treating now the switch as a carrier of one bit of information one can derive the fundamental formulas connecting thermodynamical cost of information processing with the accuracy and stability of encoding. The discussed above quantum model of a switch seems to be minimal, is exactly solvable and based on Gaussian states. Such properties usually imply that the relations obtained for this particular model are optimal and therefore can be used to draw the general conclusions (similarly to saturation of the Heisenberg uncertainty relation for Gaussian states). 
\par
The accuracy of information encoding is characterized by the error $\epsilon \in[0,1/2)$ and stability by the life-time $\tau$. Because in half of the cases of bit encoding we do nothing, in half of the cases we have to flip the spin, one can formulate, using \eqref{highT} and \eqref{kramers}, the following principles:\\

\textbf{The minimal averaged amount of work needed to encode a bit of information  is equal to}
\begin{equation}
W = \Theta \ln\frac{1}{\epsilon}. 
\label{work}
\end{equation}
\\

\textbf{The maximal life-time of encoded bit of information scales with the error in the following way}
\begin{equation}
\tau = \tau_0 \, \frac{1}{\epsilon} . 
\label{stability10}
\end{equation}
Both formulas look reasonable \footnote{An important argument supporting the formula \eqref{work} is that  its high temperature version, $W= k_BT\ln(1/\epsilon)$ has been earlier derived by Kish \cite{Kish} for a purely classical model of an electronic switch.}. One expects that encoding information in a stable carrier, say  ``engraving on a stone tablet'', should cost much more work than encoding in a less stable system like making a ``hole in a punched card``.
The formula \eqref{stability10} is based on the universaly valid Kramer's-type formula including thermal transitions and quantum tunneling. In fact it is rather obvious that:\\

\textbf{Any relation describing thermodynamical cost of information processing must take into account the accuracy and stability of information encoding, what means that the generally accepted Landauer formula $W= k_BT \ln 2$ cannot be correct for most instances of its application and interpretation  discussed in numerous papers. }\\

Another property of mechanisms which stabilize information, illustrated by the discussed switch model, is a presence of dissipation (friction) which transforms  work invested in a gate to heat. This shows a fundamental conflict between stability of information and reversibility of gates which seems to be the main obstacle for an efficient implementation of microscopically reversible computations, both classical and quantum \cite{Alicki:2012}.

\par
The formulas \eqref{work} and \eqref{stability10} can be applied to the estimation of the minimal energy $W_N$ which is necessary to perform a long computation consisting of $N$ elementary
steps \cite{Alicki:2013}. For really long computations, say $N > 10^{20}$, the leading term is very simple and given by 
\begin{equation}
W_N \simeq \Theta N\ln N,
\label{cost}
\end{equation}
which is essentially different from the estimation $k_BT N$ following from the Landauer formula\footnote{Here the other dimensionless parameters describing physical properties and confidence level are absent because they can be neglected for such a huge value of $N$. However, the additional term $\ln N$  follows from the stability and accuracy requirements.}. It is rather obvious that \eqref{cost} is more realistic, because longer computation demands more stability which cost additional energy per single operation proportional to $\ln N$. Moreover, \eqref{cost} does not vanish at $T\to 0$ what corresponds to the residual quantum fluctuations. 
\section{Quantum Szilard engine and its efficiency}
The quantum analog of the Szilard engine, proposed in \cite{Alicki:2004}, consists of a spin-1/2 with by the time-dependent Hamiltonian 
\begin{equation}
H(t)= \frac{E_{0}}{2}\bigl(f(t)^2 \hat{I} - f(t)\hat{\tau}^3 \bigr)
\label{Szil_ham}
\end{equation}
where the external control $|f(t)| \leq 1$ and $E_{0}>0$. Here, $\hat{\tau}^{\alpha}, \alpha = 1,2,3$ denote Pauli matrices of the engine's spin in order to distinquish them from the switch spin with Pauli matrices $\hat{\sigma}^{\alpha}$. The weak coupling to the heat bath at the temperature $T$ can be switched on and off.\\
The cyclic process of extracting work from a bath using measurement can be decomposed into following steps :\\

i) For the initial time $t_0$ the Hamiltonian is trivial, i.e. $f(t_0)=0$ and the spin,  coupled to the bath, is at the corresponding thermal equilibrium state $\hat{\rho}(t_0)= \frac{1}{2}\hat{I}$.\\

ii) The coupling to the bath is switched off and a measurement on the spin performed in the basis of $\hat{\tau}^3$ gives  the outcome 
$s= \pm 1$ with the corresponding projected post-measurement state $\hat{\rho }(t_1)= |s\rangle\langle s|$.\\

iii) A fast  change of the control field from the value $f(t_1)=f(t_0)=0$ to the value $f(t_2) = s$  changes the Hamiltonian to  $H(t_2) = (s E/2)(s\hat{I}-\hat{\tau}^3 )$ which increases the energy of the state $|-s\rangle$ by $E_{0}$ and  does not change the energy of
the state $|s\rangle$. As energy of the actually occupied state remains constant no work is performed during this step.
\\

iv) The spin-baths coupling is switched on and the control field is slowly reduced from the value  $f(t_2) = s$  to the value $f(t_3) = 0$ extracting work.\\
\par
One can  compute the balance of work $W(t)$ and  heat $Q(t)$ supplied to the spin and  its internal energy $E(t) $ during the full cycle $t_0\to t_1\to t_2$
using the standard definitions \cite{Alicki:1979, Alicki:2004} 
\begin{equation}
E=\mathrm{Tr }(\rho H),\quad dW = \mathrm{Tr }(\rho\, dH), \quad dQ =\mathrm{Tr } (d\rho\, H).
\label{EWQ}
\end{equation}
During the slow change of the Hamiltonian (step iv) at any moment  the spin is in the corresponding Gibbs state given by
\begin{equation}
\rho (t) = \frac{ \exp\{-\frac{H(t)}{k_B T}\}}{\mathrm{Tr} \exp\{-\frac{H(t)}{k_B T}\}},\quad t_2 \leq t\leq t_3 .
\label{Gibbs_temp}
\end{equation}
and therefore work performed by the quantum Szilard engine during the whole cycle is given by
\begin{equation}
W_{SE}(E_0) = -\int_{t_2}^{t_3}\mathrm{Tr}\bigl(\rho (t)\frac{dH(t)}{dt}\bigr) dt =  k_B T\bigl[\ln 2 - \ln\bigl(e^{-\frac{E_0}{k_B T}}+ 1\bigr)\bigr]
\label{SE_work}
\end{equation}
which for $E_0\mapsto\infty$ reaches the Landauer's value $k_B T \ln 2$. Notice that extraction of work is possible because the spin relaxes to the temporal equilibrium state on a sufficiently short time-scale.
\par
Any  measurement yields a correct result with a certain  probability $1-\epsilon$, $\epsilon\in[0, 1/2]$. Then in the step iii), with the error probability  $\epsilon$  the amount of work $E_0$ is supplied by the control field. It implies that the net extracted work is given by
\begin{equation}
W_{SE}(E_0;\epsilon) = k_B T\bigl[\ln 2 - \ln\bigl(e^{-\frac{E_0}{k_B T}}+ 1\bigr)\bigr]- \epsilon E_0.
\label{SE_work1}
\end{equation}
Maximizing $W_{SE}(E_0;\epsilon)$ with respect to $E_0$ one obtains the maximal work extracted by the quantum Szilard engine with faulty measurement
\begin{equation}
W_{SE}[\epsilon] = \max_{E_0}W_{SE}(E_0;\epsilon) = k_B T\bigl[\ln 2 - S(\epsilon) \bigr].
\label{SE_work2}
\end{equation}
where $S(\epsilon) = -\epsilon \ln\epsilon - (1-\epsilon) \ln(1-\epsilon)$ is the entropic uncertainty of the measurement. 
\par
The total work invested by the external forces depends on the $\epsilon$, i.e. an accurate measurement  costs more than a sloppy one. To estimate the cost one can use the optimal model of the switch discussed above as an element of the measuring device. Indeed, the following time-dependent Hamiltonian which couples the engine spin and the spin being a part of the switch  executes a controlled NOT gate
\begin{equation}
\hat{H}_{CNOT} = \frac{\hbar}{2} g(t) (\hat{\tau}^3 - 1)\,\hat{\sigma}^{1} .
\label{ham_measure1}
\end{equation}
Here, again, the gate field $g(t)$ is assumed to be a fast pulse concentrated around $t=0$ and satisfying $\int_{-\infty}^{\infty} g(t) dt = \pi$. 
\par
When the engine  spin is in the state ``spin up'' nothing happens to the switch. When the engine spin is ``down'' the switch oscillator changes its position.
After this ``measurement'' step the control field $f(t)$ takes the value $+1$ if nothing happens to the switch, or $-1$ if the  oscillator moves to its new position.
On the average the minimal work needed to operate this engine is equal to $W$ per cycle. 
\par
One should notice that only the change of the switch position triggers the proper action and hence there is no need to record the measurement result or to reset the switch to a standard position\footnote{There is a difference between the (true) measurement followed by writing down the measurement result on a stable information carrier and a ''measurement'' which is only a part of a feed-back control and  no information is recorded.}. \textbf{Therefore, the explanation of the engine operation need not to refer to any type of information processing.}
\par
The efficiency of the Szilard engine can be defined as the ratio of the extracted work to the minimal work $W$  needed to operate the switch and satisfies the following bound obtained numerically
\begin{equation}
\eta[\epsilon] = \frac{W_{SE}[\epsilon]}{W} = \frac{k_B T}{\Theta}\frac{\ln 2 - S(\epsilon)}{-\ln \epsilon}\leq \eta[\bar{\epsilon}]= 0.17 \frac{k_B T}{\Theta}< 0.17.
\label{SE_eff}
\end{equation}
The maximal efficiency of the Szilard engine is obtained for the value of error $\bar{\epsilon}= 0.06$.
\section{Information versus thermodynamical entropy}
The difference between the notions of information and thermodynamical entropy can be illustrated by the following example\footnote{This is a standard example used to illustrate the issues of ``Schroedinger cat states'', ``molecular structure problem'', ``transition from quantum to classical world'', etc. A sample of points of view and relevant references can be found in \cite{Joos:2003}.}. Consider two molecules: ammonia  with the chemical formula $NH_3$ and  alanine with the chemical formula $CH_3CH(NH_2)COOH$. For both molecules the lowest lying states can be described in terms of a fictitious particle in a symmetric double-well potential. The ground state $\psi_0$ and the first excited state $\psi_1$ are symmetric and  antisymmetric superpositions of the states $\psi_L , \psi_R$ localized at the left or right potential minimum
\begin{equation}
\psi_0 = \frac{1}{\sqrt{2}}\bigl(\psi_L + \psi_R\bigr), \quad \psi_1 = \frac{1}{\sqrt{2}}\bigl(\psi_L - \psi_R\bigr).
\label{double_well}
\end{equation}
For ammonia the energy difference $E_1 - E_0$ is of the order of $10^{-5} eV$ and for alanine even much lower due to a much higher energy barrier and larger effective mass of the ``particle''.
Therefore, at room temperature  ($k_B T_r = 0.0258eV$) equilibrium ensembles of both molecules can be described by the maximally mixed density matrix
\begin{equation}
\hat{\rho}_{eq} = \frac{1}{2}|\psi_0\rangle\langle \psi_0 | + \frac{1}{2}|\psi_1\rangle\langle \psi_1 |= \frac{1}{2}|\psi_L\rangle\langle|\psi_L | + \frac{1}{2}|\psi_R\rangle\langle \psi_R |
\label{double_well1}
\end{equation}
with the von Neumann entropy equal to $S(\hat{\rho}_{eq} ) = \ln 2$.\\
However, the physical meaning of the density matrix and the von Neumann entropy is completely different in both cases (compare with the discussion in \cite{Norton, Shenker}).
\par
For ammonia the equilibrium distribution is maintained by relatively fast processes of thermally induced transitions between eigenstates $\psi_0, \psi_1$ or alternatively by the tunneling between localized states $\psi_L , \psi_R$ with the inversion frequency of $23.8 GHz$. In this case $k_B S(\hat{\rho}_{eq} )= k_B\ln 2$ is the \textbf{objective physical entropy} because when ammonia is prepared in the initial pure state  $\psi_0$ or  $\psi_1$ (or alternatively, $\psi_L$ or $\psi_R$) its free energy is increased by  $k_B T ln 2$ and this amount can be transformed into work using in principle the steps iii) and iv) of the quantum Szilard engine\footnote{In fact an ammonia maser is a kind of Szilard engine. A separator splits a beam of ammonia molecules and sends the molecules in excited state into the resonant cavity where coherent microwave radiation (work) is produced.}. On the other hand both pairs of orthogonal states cannot be used as ``switch positions'' because of their instability and hence the von Neumann entropy has no information-theoretical meaning.
\par
For alanine obtained in certain chemical reactions molecules in chiral states  $\psi_L , \psi_R$ are produced with equal probabilities and both optical isomers are extremally stable with the 
inversion time of the order of $10^{29}$ years. For example, in biological context alanine appears always in left-handed form. Therefore, both states of alanine can encode information in a very stable way and the von Neumann entropy represents the \textbf{subjective lack of knowledge} about the spatial configuration of the alanine. The entropy has no thermodynamical meaning because selection of a given  left or right-handed form does not allow to extract work.
\par
More generally, for a set of $N$ disjoint and equally populated  states of the open system  at equilibrium with the bath, $k_B \ln N$ is a thermodynamical entropy if those states are ergodic with respect to the dynamics, i.e. fast dynamical transitions between any pair of them are present. Such states cannot be used to encode information. If they are not ergodic, or in other words stable with respect to dynamics, $\ln N$ describes the amount of information which can be encoded into this set of states. In the later case the knowledge which state is occupied by the system does not allow to extract any work from the heat bath.\\
\par
One can formulate a  complementarity principle for information and thermodynamical entropy :\\

\textbf{The entropy of an ensemble of quantum states can be interpreted either as thermodynamical  or information-theoretical one depending on the stability of those states with respect to dynamics.}\\

\par
It seems that the ideas of above contradict the well-established opinion that a density matrix contains all information about the system and all decomposition of  mixed states into pure or ``less mixed'' ones are equally meaningful. On the other hand there is an evidence that at least for open systems some decompositions are ``more equal''. For example, a state of a harmonic oscillator interacting with a heat bath and observed in the semi-classical regime should be rather described as a mixture of coherent states than, for example,  a mixture of energy eigenstates. Only the former are localized, structurally stable and follow classical trajectories, while the later decohere very quickly. On a more general level so-called ``pointer states'' are examples of such preferred basis \cite{Zurek:2003}. This suggests that in the very definition of a mixed state, the dynamics should play a role\footnote{Another example is the notion of statistical independence versus physical correlations. Imagine two particles on a line, one localized at origin, the other at the point $a > 0$. The classical representation of their joint state is a product of two Dirac deltas $\delta(x_1)\delta(x_2-a)$ which describes statistically independent uncorrelated systems. If the particles are connected by a rigid bar the probability distribution remains the same despite the systems are strongly correlated in the physical sense. To discover the correlations we can allow the particles to interact with the random noise with a correlation length much shorter than $a$. Then after some time the  probability distributions become very different for both cases.}.
\par
Another explanation involves the algebraic formalism of infinite quantum systems. Within this formalism there exist \emph{disjoint  states} corresponding, for example, to different thermodynamical phases. Such states are stable with respect to  local perturbations of dynamics, and macroscopically distinguishable. Their coherent superpositions do not make sense and their mixtures must be considered as ``subjective'', fully related to the lack of knowledge. In principle, one can apply these ideas to describe the difference between ammonia and alanine. Both molecules interact with electromagnetic field - a system with infinite number of degrees of freedom described by the algebraic formalism. For a small molecule the disturbance of electromagnetic field is weak and two lowest lying states of the joint systems can be approximately described by the products $\phi_1\otimes\Phi_0$ and  $\phi_2\otimes\Phi_0$
with a fixed state of the field $\Psi_0$. For large enough molecules the coupling is stronger and a kind of \emph{dynamical phase transition} appears producing two degenerated ground states  $\phi_L\otimes\Phi_L$ and $\phi_R\otimes\Phi_R$, where now the states of the field $\Phi_{L(R)}$  contain an infinite number of soft photons and are disjoint.
\par

There are interesting situations which interpolate between the fully thermodynamical and fully information-theoretical meaning of the entropy. Consider, a spin-$1/2$ particle emitted by a macroscopic ``polarizer'' which can prepare the spin degree of freedom in one of the pure states $\psi_{\mathbf{n}_j}; j =1,2,...,N$ , where ${\mathbf{n}_j}$ is a vector of the Bloch sphere determining spin's direction. The amount  of information which can be encoded in the polarizer  is equal to $\ln N$, while the amount of information which can be send using the spin degree of freedom $I_{spin}$ is bounded by the Holevo bound 
\begin{equation}
I_{spin} \leq S\bigl(\frac{1}{N}\sum_{j=1}^{N} |\psi_{\mathbf{n}_j}\rangle\langle\psi_{\mathbf{n}_j}|\bigr) \leq \ln 2.
\label{holevo}
\end{equation}
On the other hand  by connecting the position of the macroscopic polarizer with the source of the control field in the quantum Szilard engine (Section V) one can extract from the spin $k_B T\ln 2$ of work under ideal conditions\footnote{The source of this work is energy invested into  preparation of the polarized spin from the equilibrium ensemble of depolarized ones.}. As the spin free energy $F= E - TS_{ph}$, with $E=0$, is between $-k_B T\ln2$ and zero it means that the initial physical entropy of the spin was equal to zero.
Therefore, the physical entropy is not related to the information-theoretical entropies associated with this physical setting. 
\par
The above statement is correct under the assumptions that the spin states are perfectly stable.
A thermal noise acting on the spin between its state preparation and coupling to the control field  transforms any state  $|\psi_{\mathbf{n}_j}\rangle\langle\psi_{\mathbf{n}_j}|$  into $(1-2\epsilon)|\psi_{\mathbf{n}_j}\rangle\langle\psi_{\mathbf{n}_j}| + \epsilon \hat{I}$ where $\hat{I}$ is the identity matrix and $\epsilon\in[0, 1/2]$. The amount of information carried by the spin is reduced to
\begin{equation}
{I'}_{spin} \leq S\bigl(\frac{1}{N}\sum_{j=1}^{N} |\psi_{\mathbf{n}_j}\rangle\langle\psi_{\mathbf{n}_j}|\bigr) - S(\epsilon) \leq \ln 2 - S(\epsilon)
\label{holevo1}
\end{equation}
while the physical entropy increases to $k_BS(\epsilon)$,  $S(\epsilon)= -\epsilon \ln\epsilon - (1-\epsilon) \ln(1-\epsilon)$, illustrating  complementarity of  thermodynamical and information-theoretical notions.

\section{Cloning and error correction}
The main idea of ``quantum information'' and, particularly, ``quantum computations'' is to use strongly overlapping quantum states as information carriers and reversible (unitary) transformations of those states as ``gates''.
The early criticism of those ideas was concentrated on the instability of such states with respect to environmental noise. The standard remedy in a form of error correction seemed not to work because of absence of cloning - a main ingredient of classical error correction schemes. Before entering the field of ``quantum error correction'' I would like to discuss briefly those issues using a general framework introduced in \cite{Alicki:2006}.
\par

Consider a set of states (density matrices) $\mathcal{S} = \{\hat{\rho}\}$ of the quantum physical system. This set may contain all states or only a selected subset used to encode information. The overlap of two states is defined by the standard formula for the transition probability already used in \eqref{trans_prob}
\begin{equation}
(\hat{\rho}| \hat{\rho}')  = \mathrm{Tr}\Bigl(\sqrt{\sqrt{\hat\rho}\,{\hat\rho}'\sqrt{\hat\rho}}\Bigr) .
\label{trans_prob1}
\end{equation}
The dynamical maps transforming states  are always completely positive trace-preserving maps. The tensor product notation $\hat{\rho}\otimes \hat{\sigma}$
is used for product states of composite systems. One can prove the following properties :\\

A1)  The overlap  $(\hat{\rho}|\hat{\rho}')$  measures the {\sl indistinguishability} of two states  and satisfies the conditions
\begin{equation}
0  \leq    (\hat{\rho}|\hat{\rho}')  \leq 1\  ,  (\hat{\rho}|\hat{\rho}')= 1\   {\rm if\ and\ only\ if}\     \hat{\rho} =\hat{\rho}' .
\label{a1}
\end{equation}

A2) For all product states $\hat{\rho}\otimes \hat{\sigma}$, $\hat{\rho}'\otimes \hat{\sigma}'$  the following factorization holds
\begin{equation}
(\hat{\rho}\otimes \hat{\sigma} |\hat{\rho}'\otimes \hat{\sigma}') = (\hat{\rho} |\hat{\rho}')(\hat{\sigma} |\hat{\sigma}'). 
\label{a2}
\end{equation}                                      

A3) Any dynamical map $T$ does not reduce the overlap of two arbitrary states\footnote{This property can be understood as the most general manifestation of the Second Law of Thermodynamics.} 
\begin{equation}
(T\hat{\rho} | T\hat{\rho}') \geq  (\hat{\rho} |\hat{\rho}' )   .
\label{a3}
\end{equation}                                                        
\par
The above properties can be used to put a limit on \emph{self-replication} (``cloning'') process described by the following dynamical transformation 
\begin{equation}
\hat{\rho}\otimes \hat{\omega} \mapsto \hat{\rho}\otimes\hat{\rho}\otimes\hat\sigma = T(\hat\rho\otimes\hat\omega)
\label{srW}
\end{equation} 
where (using biological terminology) $\hat\omega$ is the initial state of ``food'', $\hat\sigma$ is the final state of  the "remained food",  and $T$ denotes the dynamics defined on the total system.  A simple inequality which follows from A1)-A3)
\begin{equation}
(\hat{\rho}|\hat{\rho}') =  (\hat{\rho}\otimes\hat\omega | \hat{\rho}'\otimes\hat\omega ) \leq  ( T(\hat{\rho}\otimes\hat\omega)| T(\hat{\rho}'\otimes\hat\omega) ) 
 =  (\hat{\rho}\otimes\hat{\rho}\otimes\hat\sigma |\hat{\rho}'\otimes\hat{\rho}'\otimes\hat\sigma' )    = (\hat{\rho}| \hat{\rho}')^2  (\hat\sigma |\hat\sigma')
 \leq (\hat{\rho}| \hat{\rho}')^2         
\label{ncp}
\end{equation}                                                                                     
implies that either  $\hat\rho = \hat\rho'$  or   $(\hat\rho |\hat\rho' ) = 0$, i.e. the states are either identical or \emph{disjoint}.\\
It means that:\\

 \textbf{A set of perfectly (approximately) clonable states must consist of perfectly (approximately) disjoint ones}\footnote{``Non-cloning'' of pure quantum states was first invoked in 1967 by Wigner \cite{Wig} who argued that the phenomenon of self-replication of biological molecules  and organisms contradicted the principles of quantum mechanics.  In 1982 Wooters,  \.Zurek  \cite{Woo} and   Dieks 
\cite{D} proved "no-cloning theorem"  for an arbitrary quantum unitary dynamics, (see also Ghirardi's independent proof \cite{Ghirardi}). Obviously, non-cloning holds also for overlapping classical probability distributions.}.\\

Consider now the process of recovering information (encoded in a certain set of states) which is being destroyed by the noise map $T_n$. Denoting the recovery map by $T_r$ we obtain the inequality following from A3)
\begin{equation}
(\hat{\rho}|\hat{\rho}') =  (T_r\circ T_n\hat{\rho}|T_r\circ T_n\hat{\rho}')\geq (T_n\hat{\rho}| T_n\hat{\rho}')\geq (\hat{\rho}| \hat{\rho}')
\label{correction}
\end{equation}                                                                                     
what implies that $(T_n\hat{\rho}| T_n\hat{\rho}') = (\hat{\rho}| \hat{\rho}') $, i.e. a distance between states is conserved. Hence:\\

\textbf{Information can be  directly ``recovered'' only if it is not lost at all, i.e. if the noise map acts reversibly on the chosen set of states.}\\

Self-replication process combined with recovering map can correct errors for truly irreversible noise. Namely, the new recovery condition
\begin{equation}
T_r\bigl[T_n(\hat{\rho}\otimes\hat{\rho}\otimes\hat\sigma)\bigl] = \hat{\rho}\otimes\hat{\kappa} 
\end{equation}                                                                                     
implies, instead of the inequality \eqref{correction}, the following one
\begin{equation}
(\hat{\rho}|\hat{\rho}')\geq  (\hat{\rho}|\hat{\rho}')(\hat{\kappa}|\hat{\kappa}')= (T_r\circ T_n(\hat{\rho}\otimes\hat{\rho}\otimes\hat\sigma)|T_r\circ T_n(\hat{\rho}'\otimes\hat{\rho}'\otimes\hat\sigma'))\geq (\hat{\rho}|\hat{\rho}')^2(\hat{\sigma}|\hat{\sigma}') 
\label{correction1}
\end{equation}                                                                                     
which does not introduce any  additional restrictions on the set of states $\{\rho\}$ or the noise map.
\par
Those elementary results show that:\\

\textbf{A true error correction is possible only for sets of disjoint states admitting also self-replication.}\\

The so-called ``quantum error correction'' \cite{QEC} is in fact ``error prevention'' realized by suitable ``decoherence-free'' subsystems, i.e. quantum degrees of freedom which are essentially not influenced by noise. Beside the decoherence-free subsystems generated by some symmetries which are always approximate and non-scalable, the other examples from the vast literature are usually extremely complicated and based on semi-phenomenological error models. 
\par
In my opinion the complexity and semi-phenomenological character of the existing fault-tolerance schemes permits to hide there ``Maxwell demons'' which  are otherwise not allowed within a complete quantum Hamiltonian description. For example, in the standard approach  based on ``threshold theorems'' a mechanism involving ``fresh qubits'' is crucial, but  was never scrutinized within a Hamiltonian model ( other flaws in the physical assumptions are discussed in \cite{Alicki:QEC}). Another example of a  different but equivalent scheme of computation, discussed in \cite{emerson}, employs as a basic resource ``magic states'' with negative quasi-probability functions. On the other hand it is known that such ``truly quantum'' states are particularly sensitive to environmental noise.

\section{Concluding remarks}
The view that information is an abstract entity independent of the physical nature of a carrier, which differs from the thermodynamical entropy and should not be included in the physical energy-entropy balance, does not mean that physics is irrelevant for the field of information processing. On the contrary, the smaller and hence less stable are devices used in the information technology the more important are physical questions concerning the nature of noise, mechanisms of stability, life-times of information carrying states, an obvious conflict between stability and reversibility of gates, thermodynamical costs of computing etc.  Some of these questions are discussed in this essay  using new relations derived from a reasonable model which should give more realistic estimations and bounds. Those bounds take into account  accuracy and stability of information processing with respect to environmental noise. They are different from the generally accepted ones which are  based on the equivalence of information and physical entropy advocated by Szilard, Brillouin,  Landauer,  Bennett and others.

\textbf{Acknowledgements}\\
The author acknowledges the support by the FNP TEAM project cofinanced by EU Regional Development Fund.\\

\end{document}